\documentclass[aps,pre,showpacs,preprintnumbers,amsmath,amssymb]
{revtex4}

\usepackage{graphicx}

\begin{document}

\newcommand\etal{\mbox{\textit{et al.}}}

\title{Renormalization-group and numerical analysis of a noisy Kuramoto-Sivashinsky equation in 1+1 dimensions}

\author{K. Ueno$^{1}$}

\altaffiliation[Present address:]
{Department of Computational Science and Engineering, 
Graduate School of Engineering, Nagoya University, 
Chikusa, Nagoya 464-8601, Japan}

\email{ueno@fcs.coe.nagoya-u.ac.jp}

\author{H. Sakaguchi$^{2}$}

\author{M. Okamura$^{1}$}

\affiliation{$^{1}$Research Institute for Applied Mechanics, Kyushu University, Kasuga, Fukuoka 816-8580, Japan \\
$^{2}$Department of Applied Science for Electronics and Materials, Interdisciplinary Graduate School of Engineering Sciences, Kyushu University, Kasuga, Fukuoka 816-8580, Japan}


\begin{abstract}
The long-wavelength properties of a noisy Kuramoto-Sivashinsky (KS) equation in 1+1 dimensions are investigated by use of the dynamic renormalization group (RG) and direct numerical simulations. It is shown that the noisy KS equation is in the same universality class as the Kardar-Parisi-Zhang (KPZ) equation in the sense that they have scale invariant solutions with the same scaling exponents in the long-wavelength limit. The RG analysis reveals that the RG flow for the parameters of the noisy KS equation rapidly approach the KPZ fixed point with increasing the strength of the noise. This is supplemented by the numerical simulations of the KS equation with a stochastic noise, in which the scaling behavior of the KPZ equation can be easily observed even in the moderate system size and time.
\end{abstract}


\maketitle

\section{introduction}
The most attractive models for surface roughening are the KPZ \cite{Kardar} and the KS \cite{Kuramoto, Sivashinsky} equations. The KPZ equation has a positive surface tension coefficient and is driven by a random forcing, while the KS equation is completely deterministic and is driven by inherent instabilities caused by a negative surface tension coefficient. In spite of such a difference between  these equations, it was conjectured by Yakhot that the large-scale properties of the KS equation in 1+1 dimensions are described by the KPZ equation (a noisy Burgers equation) \cite{Yakhot,Forster,Yakhot-She}. 

In order to clarify the claim, a number of numerical investigations in 1+1 dimensions have been developed \cite{Zaleski, Sneppen, Hayot,Sakaguchi}. The present understanding for this conjecture is that the spatiotemporal chaos generated by the negative surface tension becomes renormalized at long-wavelength into an effective positive surface tension and an effective noise term. In practice, the parameters in the effective stochastic equation that describes the long-wavelength properties of the KS equation in 1+1 dimensions were determined by the coarse-graining method \cite{Zaleski, Hayot} and other methods \cite{Sneppen, Sakaguchi}, and on large scales it was shown that the KS equation behaves like the KPZ equation. However, Zaleski $\etal$ were unable to see the dynamic scaling phenomena which are numerically more costly to obtain precisely \cite{Zaleski, Hayot, Sakaguchi}. Sneppen $\etal$ observed the onset of crossover to asymptotic KPZ scaling by extensive numerical simulations on large system \cite{Sneppen}, but they could not find the KPZ dynamic scaling clearly.

In the theoretical point of view, Yakhot used a RG approach \cite{Yakhot}. Unfortunately, the theoretical assertion was based on uncertain ground, since the KS equation was treated with a perturbation theory around the unstable propagator which gives rise to an uncontrolled divergence. On the other hand, L'vov $\etal$ proved the identity of the scaling behavior of the one-dimensional KPZ and KS equations in the long-wavelength limit under the locality of nonlinear interaction of these equations in the wave number space \cite{L'vov}. They developed a perturbative treatment around the renormalized rather than the bare propagator by assuming the existence of such a renormalized propagator from the first. However, this approach needs to find a self-consistent scheme to examine its properties \cite{Procaccia}.

There are other attempts to answer the claim addressed above, in which whether the KS equation with a stochastic noise term and the KPZ equation fall into the same universality class in 1+1 dimensions was investigated numerically \cite{Cuerno, Karma} and theoretically \cite{R-Cuerno}.
One is the numerical study of dynamic roughening in surfaces eroded by ion sputtering. The early and late time dynamics of an erosion model, which is inherently stochastic,  were found to be the same as those obtained from the noisy KS equation \cite{Cuerno}. 
The other is the numerical simulation of a nonlinear stochastic equation describing the meandering of an isolated step on a crystal face grown from vapor. The nonlinear stochastic equation takes the form of the noisy KS equation above a critical supersaturation. The roughening function characterizing the step roughness obtained from the KS equation without a noise term coincides with that obtained from the one with a noise term, with increasing distance from equilibrium (see FIG. 1 in \cite{Karma}). This result suggests that the noisy KS equation exhibits the same behavior as the KS equation. 
Moreover, it is reported in Refs. \cite{Cuerno, Karma} that the steady-state spectrum $\langle h_{k}h_{-k} \rangle$ for the height $h$ of a growing interface determined from the noisy KS equation was found to obey the generic $k^{-2}$ scaling of the KPZ and deterministic KS equations for small wave number $k$, but the dynamic scaling properties were not clearly found in \cite{Cuerno} or not investigated in \cite{Karma}. This may be due to the smallness of the noise strength or limitation of the system size and time of the numerical simulations. 
In order to clarify the relation between the KS and KPZ equations from a different theoretical point of view, a dynamic RG analysis was performed for the KS equation with a nonconserved noise, and a stable fixed point of the RG flow equations for the parameters of the noisy KS equation was found, which was identified as the KPZ fixed point in Ref. \cite{R-Cuerno}. However, their values of the roughness and dynamic scaling exponents in 1+1 dimensions are different from those at the exact KPZ fixed point. 

The only difference between such a noisy version of the KS equation and the KS system is the effective noise term which originates from both deterministic noise (i.e., chaos) and stochastic noise \cite{Procaccia}. It is important to understand how the interplay between these two noises determines the roughness of the surface. 
In this paper, we apply the RG analysis to the KS equation with conserved and nonconserved noises to improve the RG results in Ref. \cite{R-Cuerno}. From the prediction of the RG results, we find a way to circumvent the limitation of the system size and time in the numerical simulations for the deterministic KS equation. 

The outline of this paper is as follows. In Sec. II, we perform the RG analysis for the noisy KS equation. In Sec. III, a part of the results predicted by the RG analysis is confirmed by the numerical simulations of the KS equation with a stochastic noise term. The conclusion is given in Sec. IV.

\section{renormalization-group analysis}
\subsection{RG flow equations}
The noisy KS equation in one dimension is
\begin{equation}
h_{t}=\nu h_{xx}-Kh_{xxxx}+\frac{\lambda}{2}(h_{x})^{2}+\eta(x,t),
\label{eq:intro1}
\end{equation}
where the subscripts denote partial derivatives. Here $h(x,t)$ describes the height profile of a one-dimensional surface above a substrate point $x$ at time $t$, $\nu$ is a negative surface tension coefficient, $K$ is a positive surface diffusion coefficient, and $\lambda$ is the strength of the nonlinearity. $\eta(x,t)$ is a Gaussian white noise with zero mean and the correlation
\begin{equation}
\langle\eta(x,t)\eta(x',t')\rangle=\left[2D-2D_{d}\frac{\partial^{2}}{\partial x^{2}}\right]
\delta(x-x')\delta(t-t').
\label{eq:intro2}
\end{equation}
Here $\eta$ is composed of the nonconserved and conserved noises whose strength is $D$ and $D_{d}$, respectively \cite{Barabasi}. 
The $D_{d}$ term in Eq. (\ref{eq:intro2}) was not taken into account in Ref. \cite{R-Cuerno} when performing the RG calculations. By introducing this term, we will obtain more reasonable RG results as stated in detail below.
An equivalent equation to Eq. (\ref{eq:intro1}) is obtained for $u=-h_{x}$ and $f=-\eta_{x}$ as
\begin{equation}
u_{t}=\nu u_{xx}-Ku_{xxxx}-\lambda u u_{x}+f(x,t).
\label{eq:intro3}
\end{equation}
When $\lambda=1$, the nonlinear term in Eq. (\ref{eq:intro3}) is the same as that in the one-dimensional analog of the Navier-Stokes equation, then the variable $u(x,t)$ can be interpreted as a one-dimensional velocity field in a compressible fluid \cite{Forster, Yakhot-She}. The dynamic RG can be described through the Fourier modes with wave number $k$ and frequency $\omega$, in terms of which Eq. (\ref{eq:intro3}) takes the form
\begin{equation}
(-i\omega+\nu k^{2}+Kk^{4})u(k, \omega)=f(k, \omega)-\frac{i\lambda}{2}k\int_{|p| \leq \Lambda_{0}}\frac{dp}{2\pi}\int_{-\infty}^{\infty}\frac{d\Omega}{2\pi}
 u(p, \Omega)u(k-p, \omega-\Omega),
\label{eq:Fs1}
\end{equation}
where $\Lambda_{0} \equiv \pi/\Delta x$ is an upper cutoff in Fourier space, $\Delta x=L/N$ is the lattice spacing in the real space, $L$ and $N$ being the system size and the number of grid, respectively. It follows from Eq. (\ref{eq:intro2}) that the Fourier transform of $f(x,t)$ satisfies
\begin{equation}
\langle f(k, \omega)f(k', \omega')\rangle=2(2\pi)^{2}k^{2}(D+D_{d}k^{2})\delta(k+k')\delta(\omega+\omega').
\label{eq:Fs2}
\end{equation}

First, following to the RG procedure \cite{Kardar, R-Cuerno, Barabasi, Golubovic, Forster}, we divide the velocity $u(k, \omega)$ into two components $u^{>}(k, \omega)$ and $u^{<}(k, \omega)$, with the wave number satisfying $\Lambda(l)\equiv\Lambda_{0} e^{-l} \leq |k| \leq \Lambda_{0}$ and $|k| \leq \Lambda(l)$, $l$ being a parameter. We eliminate (i.e., integrate away) the ``fast" modes $u^{>}(k, \omega)$, leading to an equation for the ``slow" modes $u^{<}(k, \omega)$ given by
\begin{eqnarray}
\left[-i\omega+\nu k^{2}+Kk^{4}+\Sigma(k, \omega)\right]u^{<}(k, \omega)
&=&f^{<}(k, \omega)
-\frac{i\lambda}{2}k\int_{|p| \leq \Lambda(l)}
\frac{dp}{2\pi}\int_{-\infty}^{\infty}\frac{d\Omega}{2\pi}
\nonumber \\
& & 
\times u^{<}(p, \Omega)u^{<}(k-p, \omega-\Omega),
\label{eq:Fs3}
\end{eqnarray}
and $f^{<}(k, \omega)$ satisfies
\begin{equation}
\langle f^{<}(k, \omega)f^{<}(k', \omega')\rangle
=2(2\pi)^{2}k^{2}\left[D+D_{d}k^{2}+\Phi(k, \omega)\right]
\delta(k+k')\delta(\omega+\omega').
\label{eq:Fs4}
\end{equation}
$\Sigma(k, \omega)$ in Eq. (\ref{eq:Fs3}) and $\Phi(k, \omega)$ in Eq. (\ref{eq:Fs4}) are given in the one-loop approximation as
\begin{eqnarray}
\Sigma(k, \omega)&=&\lambda^{2}k\int_{\Lambda(l) \leq |p| \leq \Lambda_{0}}
\frac{dp}{2\pi}\int_{-\infty}^{\infty}\frac{d\Omega}{2\pi}
\left[|G_{0}(p, \Omega)|^{2}G_{0}(k-p, \omega-\Omega)p^{2}(k-p)(D+D_{d}p^{2}) \right.\nonumber \\
&& \left.
+|G_{0}(k-p, \omega-\Omega)|^{2}G_{0}(p, \Omega)(k-p)^{2}p\left(D+D_{d}(k-p)^{2}\right) \right],
\label{eq:Fs5}
\end{eqnarray}
and 
\begin{eqnarray}
\Phi(k, \omega)&=&
\lambda^{2}\int_{\Lambda(l) \leq |p| \leq \Lambda_{0}}
\frac{dp}{2\pi}\int_{-\infty}^{\infty}\frac{d\Omega}{2\pi}
|G_{0}(p, \Omega)|^{2}
|G_{0}(k-p, \omega-\Omega)|^{2} \nonumber \\
& &\times p^{2}(k-p)^{2}
(D+D_{d}p^{2})\left(D+D_{d}(k-p)^{2}\right),
\label{eq:Fs6}
\end{eqnarray}
where $G_{0}(k,\omega)=1/(-i\omega+\nu k^{2}+Kk^{4})$ is the bare propagator.

We are interested in the scaling behavior in the large system size and long time region, which correspond to $k \rightarrow 0$ and $\omega \rightarrow 0$ limits, respectively. We first integrate the right hand sides of Eqs. (\ref{eq:Fs5}) and (\ref{eq:Fs6}) over $\Omega$. Next, setting $\omega=0$ and expanding the integrands in Eqs. (\ref{eq:Fs5}) and (\ref{eq:Fs6}) into Taylor's series up to order $k^{4}$ and $k^{2}$, respectively, we can express them as $\Sigma=\delta\nu k^{2}+\delta Kk^{4}$ and $\Phi=\delta D+\delta D_{d}k^{2}$. 
Since the linear part of Eq. (\ref{eq:intro3}) becomes unstable for $\nu<0$, the bare propagator $G_{0}(k,\omega)$ has a pole for the zero frequency at the wave number $k=k_{0}=(|\nu|/K)^{1/2}$ if $\Lambda(l)<k_{0}$. 
In order to avoid such an uncontrolled divergence, Procaccia $\etal$ assumed the renormalized propagator and correlation function from the first, since there are rigorous proofs for the existence and boundedness of the solutions of the KS equation \cite{L'vov, Procaccia}. Hence the propagator and correlation function in their $\Sigma(k, \omega)$ and $\Phi(k, \omega)$ are expressed in terms of the renormalized ones from the first. 
On the other hand in the RG method, if the integrations are performed over an infinitesimal wave number shell $\Lambda_{0} (1-\delta l) \leq |k| \leq \Lambda_{0}$ only, one can avoid the divergence due to the singularity of the bare propagator \cite{R-Cuerno}. Then all calculations of $\delta\nu$, $\delta K$, $\delta D$, and $\delta D_{d}$ can be evaluated up to the first order $\delta l$ without any problem, and we can define the renormalized parameters, $\nu^{<} \equiv \nu+\delta\nu$, $K^{<} \equiv K+\delta K$, $D^{<} \equiv D+\delta D$, and $D_{d}^{<} \equiv D_{d}+\delta D_{d}$. There is no correction to $\lambda$, which is a consequence of the Galilean invariance \cite{Forster}, therefore, $\lambda^{<} \equiv \lambda$.

Second, in the RG procedure, we perform the rescaling, $\tilde{k}=(1+\delta l)k$, $\tilde{\omega}=(1+z\delta l)\omega$, and $\tilde{u}(\tilde{k}, \tilde{\omega})=[1-(\alpha+z)\delta l]u^{<}(k, \omega)$, where $\alpha$ and $z$ are the roughness and dynamic scaling exponents, respectively. The combined transformation yields the renormalized and rescaled parameters, $\tilde{\nu}=[1+(z-2)\delta l]\nu^{<}$, $\tilde{K}=[1+(z-4)\delta l]K^{<}$, $\tilde{\lambda}=[1+(\alpha+z-2)\delta l]\lambda^{<}$, $\tilde{D}=[1+(z-2\alpha-1)\delta l]D^{<}$, and $\tilde{D}_{d}=[1+(z-2\alpha-3)\delta l]D_{d}^{<}$, where the variables with tilde denote the rescaled ones, while the variables without tilde are ones in the original scale. In the limit $\delta l \rightarrow 0$, we obtain the one-loop RG flow equations describing the change in the parameters of the noisy KS equation under the RG transformation:
\begin{equation}
\frac{d\tilde{\nu}}{dl}=\tilde{\nu}\left[z-2+\frac{G}{F(1+F)^{3}}\left\{3+F+(1-F)\frac{H}{G}\right\}\right],
\label{subeq:nu}
\end{equation}

\begin{equation}
\frac{d\tilde{K}}{dl}=
\tilde{K}\left[z-4+\frac{G}{2(1+F)^{5}}\left\{26-F+2F^{2}+F^{3}
+(2-21F+6F^{2}+F^{3})\frac{H}{G}\right\}
\right],
\label{subeq:K}
\end{equation}

\begin{equation}
\frac{d\tilde{\lambda}}{dl}=\tilde{\lambda}[\alpha+z-2],
\label{subeq:lambda}
\end{equation}

\begin{equation}
\frac{d\tilde{D}}{dl}=
\tilde{D}\left[z-2\alpha-1+\frac{G}{(1+F)^{3}}\left(1+\frac{H}{G}\right)^{2}\right],
\label{subeq:D}
\end{equation}

\begin{eqnarray}
\frac{d\tilde{D}_{d}}{dl}&=&
\tilde{D}_{d}\left[z-2\alpha-3+\frac{G^{2}}{2H(1+F)^{5}}
\right. \nonumber \\
& & \left. 
\times \left\{16+3F+F^{2}+2(9-5F)\frac{H}{G}
+(2-13F-F^{2})\left(\frac{H}{G}\right)^{2}\right\}
\right],
\label{subeq:Dd}
\end{eqnarray}
where we have defined the dimensionless coupling constants $F(l)=\tilde{\nu}(l)/\left(\tilde{K}(l)\Lambda_{0}^{2}\right)$, $G(l)=\tilde{\lambda}(l)^{2}\tilde{D}(l)/\left(4\pi \tilde{K}(l)^{3}\Lambda_{0}^{7}\right)$, and $H(l)=\tilde{\lambda}(l)^{2}\tilde{D}_{d}(l)/\left(4\pi \tilde{K}(l)^{3}\Lambda_{0}^{5}\right)$, which are expressed in terms of the rescaled variables. From Eqs. (\ref{subeq:nu})-(\ref{subeq:Dd}), we can obtain the flow equations for $F$, $G$, and $H$:
\begin{equation}
\frac{dF}{dl}=2F+\frac{G}{2(1+F)^5}\left\{6-12F+11F^{2}-F^{4}
+\left(2+19F^{2}-8F^{3}-F^{4}\right)\frac{H}{G}\right\},
\label{subeq:F}
\end{equation}

\begin{eqnarray}
\frac{dG}{dl}&=&
7G-\frac{G^{2}}{2(1+F)^5}\left\{76-7F+4F^{2}+3F^{3}
+\left(2-71F+14F^{2}+3F^{3}\right)
\frac{H}{G}\right. \nonumber \\
& & \left.
-2(1+F)^{2}\left(\frac{H}{G}\right)^{2}\right\},
\label{subeq:G}
\end{eqnarray}

\begin{eqnarray}
\frac{dH}{dl}&=&
5H+\frac{G^{2}}{2(1+F)^5}\left\{16+3F+F^{2}-\left(60+7F+6F^{2}+3F^{3}\right)\frac{H}{G}
\right. \nonumber \\
& & \left. 
-\left(4-50F+19F^{2}+3F^{3}\right)\left(\frac{H}{G}\right)^{2}\right\}.
\label{subeq:H}
\end{eqnarray}
There is another expression for the dimensionless coupling constants, $f(l)=\tilde{K}(l)\Lambda_{0}^{2}/\tilde{\nu}(l)$, $g(l)=\tilde{\lambda}(l)^{2}\tilde{D}(l)/\left(4\pi \tilde{\nu}(l)^{3}\Lambda_{0}\right)$, and $h(l)=\tilde{\lambda}(l)^{2}\tilde{D}_{d}(l)\Lambda_{0}/\left(4\pi \tilde{\nu}(l)^{3}\right)$.  However, they are not convenient when $\tilde{\nu}$ is flowing towards zero, at which $f(l)$, $g(l)$, and $h(l)$ diverge \cite{R-Cuerno}. $F(l)$, $G(l)$, and $H(l)$ can be expressed as $F(l)=1/f(l)$, $G(l)=g(l)/f(l)^{3}$, and $H(l)=h(l)/f(l)^{3}$.

\subsection{Fixed point of RG flow equations}

When putting $\tilde{D}_{d}=0$ (i.e., $H=0$ ) in Eqs. (\ref{subeq:F}) and (\ref{subeq:G}), they do not reduce to the corresponding equations (20) and (19) at $d=1$ in \cite{R-Cuerno}. Therefore, we need to make some comments on the results in \cite{R-Cuerno}, in which Cuerno $\etal$ carried out the RG calculations for Eq. (\ref{eq:intro1}) by taking into account only $D$ term in Eq. (\ref{eq:intro2}), and found a stable fixed point at $(F^{*},G^{*})=(-25.25, -722.8)$ at $d=1$ with exponents $z=1.46$, $\alpha=0.54$, and thus $\beta=\alpha/z=0.37$. They identified this as the KPZ fixed point. Their RG flow equations in the limit of $|\tilde{\nu}| \ll \tilde{K}\Lambda_{0}^{2}$ reduce to Eq. (7) in \cite{Golubovic}. However, the negative values of $F^{*}$ and $G^{*}$ mean that unstable modes appear above the wave number $\tilde{k}=(\tilde{\nu}^{*}/|\tilde{K}^{*}|)^{1/2}$ because $\tilde{\nu}^{*}>0$, $\tilde{K}^{*}<0$ at the fixed point. Moreover, the substitution of $\alpha=0.54$ in the scaling function of the energy spectrum yields $E(k) \propto k^{-0.08}$ for $k \rightarrow 0$, then we cannot say definitely that the energy spectrum for $k \rightarrow 0$ is independent of $k$ (see the scaling solution of the energy spectrum and the fluctuation-dissipation theorem below). 

We have checked that the results at $d=1$ in \cite{R-Cuerno} are obtained by making the substitutions $p \rightarrow p+k/2$ and $\Omega \rightarrow \Omega+\omega/2$ in Eqs. (\ref{eq:Fs5}) and (\ref{eq:Fs6}), but without changing the integral region from $\Lambda(l) \leq |p| \leq \Lambda_{0}$ to $\Lambda(l) \leq |p+k/2| \leq \Lambda_{0}$. We have performed the same integrations by replacing the integral region. As a result, we have obtained the same RG flow equations for $\tilde{\nu}$ and $\tilde{D}$ as Ref. \cite{R-Cuerno}, but a different RG flow equation for $\tilde{K}$, since the effect of the replacement of the integral region appears when expanding the results up to order $k^{4}$. We have carried out the same calculations of Eqs. (\ref{eq:Fs5}) and (\ref{eq:Fs6}) by directly integrating them without making the substitutions $p \rightarrow p+k/2$ and $\Omega \rightarrow \Omega+\omega/2$, and confirmed to lead to the same results as those obtained by the replacement of both the variables and range of integration. The results are Eqs. (\ref{subeq:F}) and (\ref{subeq:G}) without $H$ term. Then we found a stable fixed point at $(F^{*},G^{*})=(13.1868, 1064.43)$ with exponents $z=1.54$, $\alpha=0.46$, and thus $\beta=0.30$. We can recover the positive value of $F^{*}$ (i.e., $\tilde{K}^{*}>0$), therefore, it is found that the problem of the negative value of $\tilde{K}$ at the fixed point is not attributed to the one-loop approximation \cite{R-Cuerno}. Even if we made such a modification, the values of the exponents are still different from the exact values of the KPZ fixed point. 

In order to overcome this problem we have introduced $D_{d}$ term in the noise correlator like Eq. (\ref{eq:intro2}). As a result, the parameter space $(F, G)$ is extended to $(F, G, H)$, and a stable fixed point for the RG flow equations (\ref{subeq:F})-(\ref{subeq:H}) is found at $(F^{*}, G^{*}, H^{*})=(10.7593, 680.652, 63.2614)$ with the scaling exponents, $z=1.5$, $\alpha=0.5$, and thus $\beta=1/3$. Here the values of $z$ and $\alpha$ are determined from equations $d\tilde{\nu}/dl=0$ and $d\tilde{D}/dl=0$ by the use of values of $(F^{*}, G^{*}, H^{*})$. The scaling exponents are exactly those at the KPZ fixed point. 
Due to the Galilean invariance of Eq. (\ref{eq:intro3}), $\lambda$ should remain unchanged under rescaling, therefore, Eq. (\ref{subeq:lambda}) leads to the scaling relation $\alpha+z=2$. Indeed, the values of $z$ and $\alpha$ obtained above satisfy this scaling identity. 

Figure \ref{fig:F-G,F-H} displays the trajectories of the dimensionless coupling constants $(F(l), G(l), H(l))$ under the RG transformation for an initial region $0 \leq l \leq 2.1$ with respect to four initial values $\left(-1/\Lambda_{0}^{2}, 0, 0\right)$, $\left(-1/\Lambda_{0}^{2}, 0.1/(4\pi \Lambda_{0}^{7}), 0\right)$, $\left(-1/\Lambda_{0}^{2}, 5/(4\pi \Lambda_{0}^{7}), 0\right)$, and $\left(-1/\Lambda_{0}^{2}, 40/(4\pi \Lambda_{0}^{7}), 0\right)$, where we have set $\tilde{\nu}(0)=-1$, $\tilde{K}(0)=1$, $\tilde{\lambda}(0)=1$, $\tilde{D}(0)=D$ $(D=0, 0.1, 5, 40)$, $\tilde{D}_{d}(0)=0$, and $\Lambda_{0}=\pi/\Delta x=\pi/0.5$. These initial values will be used in the numerical simulations in the next section. Except for the case of $\left(-1/\Lambda_{0}^{2}, 0, 0\right)$ (i.e., $\tilde{D}(0)=\tilde{D}_{d}(0)=0$), the remaining three trajectories reach the same fixed point $(F^{*}, G^{*}, H^{*})=(10.7593, 680.652, 63.2614)$. Moreover, it is found that the RG flow for $(F(l), G(l), H(l))$ rapidly approaches the KPZ fixed point with increasing the strength of $D$. The change in sign of $F$ from $F<0$ to $F>0$ indicates that $\tilde{\nu}$ changes from a negative value to a positive one. Once the value of $D$ takes a different value from zero, the RG trajectory deviates from the $F$ axis. For much smaller $D$ the RG trajectories go along the $F$ axis for much longer time (if we regard the parameter $l$ as time), and then they turn the direction to the right and finally reach the KPZ fixed point. Therefore for very small $D$ it takes much time (i.e., many RG transformations) to reach the KPZ fixed point. This behavior is also the case for $\tilde{D}(0)=0$ and $\tilde{D}_{d}(0)\neq 0$. 
We will numerically confirm in the next section that the KPZ scaling can be more easily observed by adding larger noises to the KS equation.
In the case of $\tilde{D}(0)=\tilde{D}_{d}(0)=0$, the initial value $(-1/\Lambda_{0}^{2}, 0, 0)$ goes to $(-\infty, 0, 0)$. If we see this behavior in the parameter space $(f(l), g(l), h(l))$, the initial value $(-\Lambda_{0}^{2}, 0, 0)$ approaches the fixed point $(0,0,0)$.

\begin{figure}
\begin{center}
\includegraphics[width=6cm,height=6cm,keepaspectratio,clip]{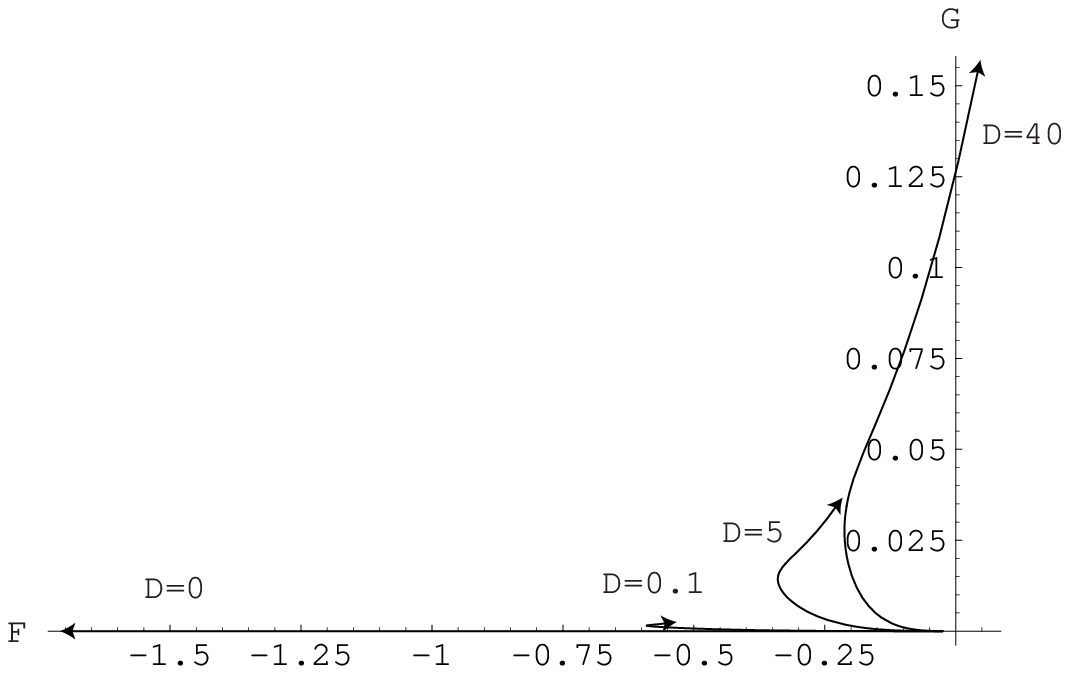}
\includegraphics[width=6cm,height=6cm,keepaspectratio,clip]{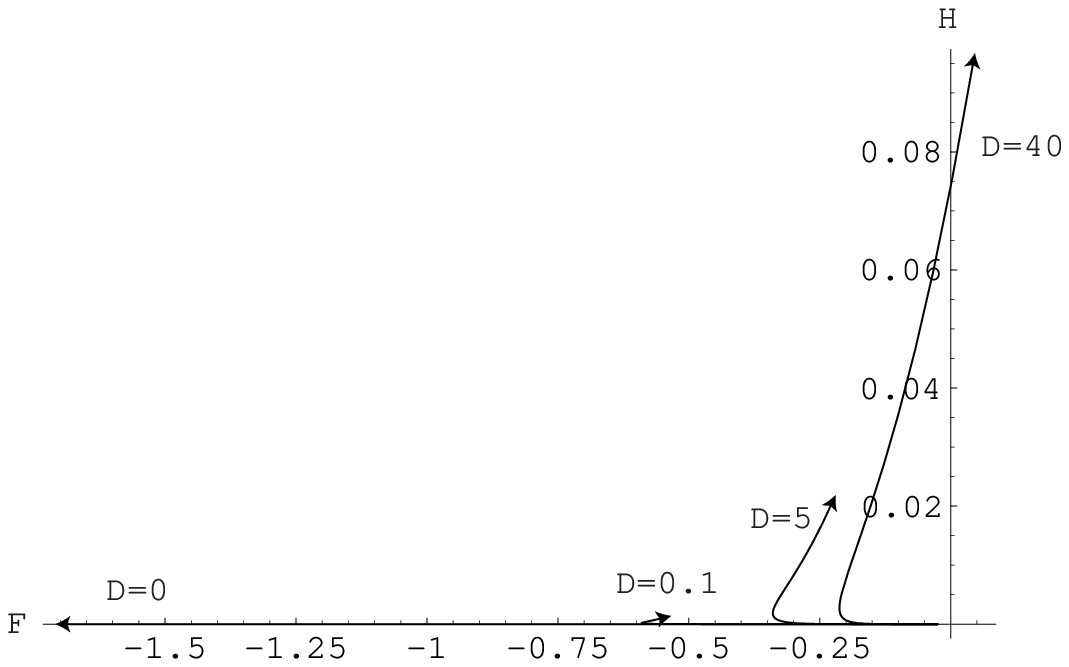}
\end{center}
\caption{The RG trajectories in the parameter space $(F, G, H)$ for $0 \leq l \leq 2.1$, projected on (a) the $(F, G)$ and (b) the $(F, H)$ plane, for initial values $\tilde{\nu}(0)=-1$, $\tilde{K}(0)=1$, $\tilde{\lambda}=1$, $\tilde{D}(0)=D$ $(D=0, 0.1, 5, 40)$, and $\tilde{D}_{d}(0)=0$.}
\label{fig:F-G,F-H}
\end{figure}

\subsection{Undoing the rescaling}

Our real interest is in the effective theories without rescaling, so that, for comparison of the RG results with the numerical simulations in the next section, one must ``undo" the rescalings \cite{Eyink}. In order to return to the original scale, we have to make the following transformations such that
$k=e^{-l}\tilde{k}$, 
$\omega=e^{-zl}\tilde{\omega}$, 
$\nu(l)=e^{-(z-2)l}\tilde{\nu}(l)$,
$K(l)=e^{-(z-4)l}\tilde{K}(l)$,
$\lambda(l)=e^{-(\alpha+z-2)l}\tilde{\lambda}(l)$,
$D(l)=e^{-(z-2\alpha-1)l}\tilde{D}(l)$,
and
$D_{d}(l)=e^{-(z-2\alpha-3)l}\tilde{D}_{d}(l)$.
In terms of the variables in the original scale, the coupling constants can be expressed as $F(l)=\nu(l)/\left(K(l)\Lambda(l)^{2}\right)$, $G(l)=\lambda(l)^{2}D(l)/\left(4\pi K(l)^{3}\Lambda(l)^{7}\right)$, and $H(l)=\lambda(l)^{2}D_{d}(l)/\left(4\pi K(l)^{3}\Lambda(l)^{5}\right)$, where we have used $\Lambda(l)=\Lambda_{0}e^{-l}$. Under the RG transformation the cutoff $\Lambda_{0}$ is fixed, while the cutoff $\Lambda(l)$ in the original scale corresponds to an appropriately chosen cutoff introduced in \cite{Zaleski,Hayot}, in which short-wavelength degrees of freedom $u(k)$ with $|k|>\Lambda(l)$ were explicitly eliminated. Going back to the original scale, the flow equations (\ref{subeq:F})-(\ref{subeq:H}) remain unchanged, but the flow equations (\ref{subeq:nu})-(\ref{subeq:Dd}) are changed to
\begin{equation}
\frac{d\nu}{dl}=\nu\left[\frac{G}{F(1+F)^{3}}\left\{3+F+(1-F)\frac{H}{G}\right\}\right],
\label{subeq:orig-nu}
\end{equation}

\begin{equation}
\frac{dK}{dl}=
K\left[\frac{G}{2(1+F)^{5}}\left\{26-F+2F^{2}+F^{3}
+(2-21F+6F^{2}+F^{3})\frac{H}{G}\right\}
\right],
\label{subeq:orig-K}
\end{equation}

\begin{equation}
\frac{d\lambda}{dl}=0,
\label{subeq:orig-lambda}
\end{equation}

\begin{equation}
\frac{dD}{dl}=
D\left[\frac{G}{(1+F)^{3}}\left(1+\frac{H}{G}\right)^{2}\right],
\label{subeq:orig-D}
\end{equation}

\begin{eqnarray}
\frac{dD_{d}}{dl}&=&
D_{d}\left[\frac{G^{2}}{2H(1+F)^{5}}
\right. \nonumber \\
& & \left. 
\times \left\{16+3F+F^{2}+2(9-5F)\frac{H}{G}
+(2-13F-F^{2})\left(\frac{H}{G}\right)^{2}\right\}
\right].
\label{subeq:orig-Dd}
\end{eqnarray}
These flow equations take the similar forms as Eqs. (\ref{subeq:nu})-(\ref{subeq:Dd}), but in the original scale it is not necessary to consider the contribution from the rescalings. 

\begin{figure}
\begin{center}
\includegraphics[width=6cm,height=6cm,keepaspectratio,clip]{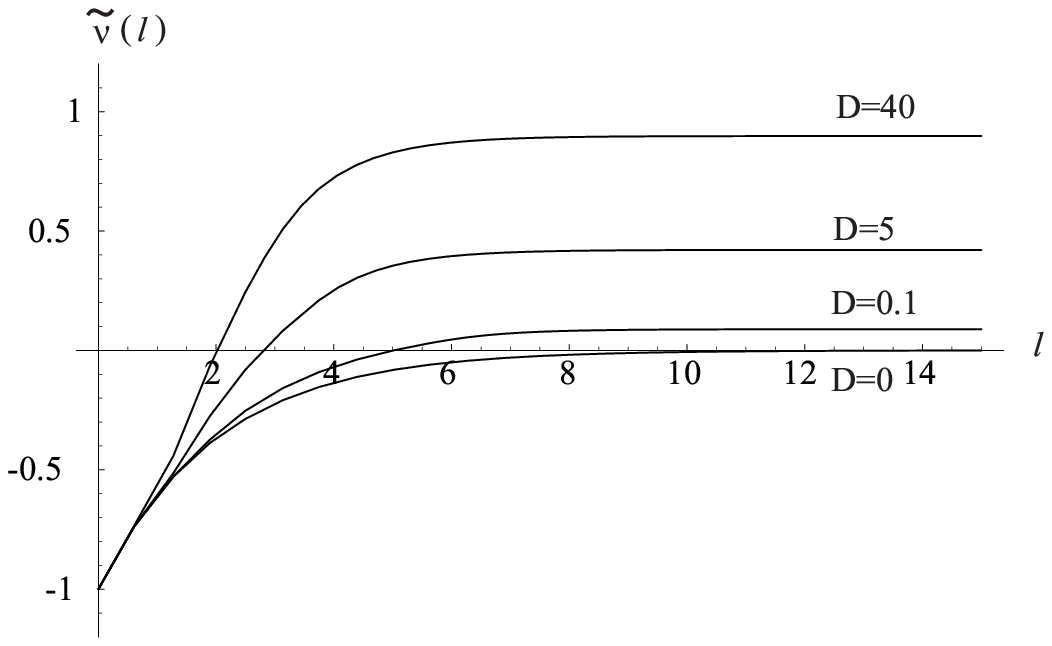}
\hspace{5mm}
\includegraphics[width=6cm,height=6cm,keepaspectratio,clip]{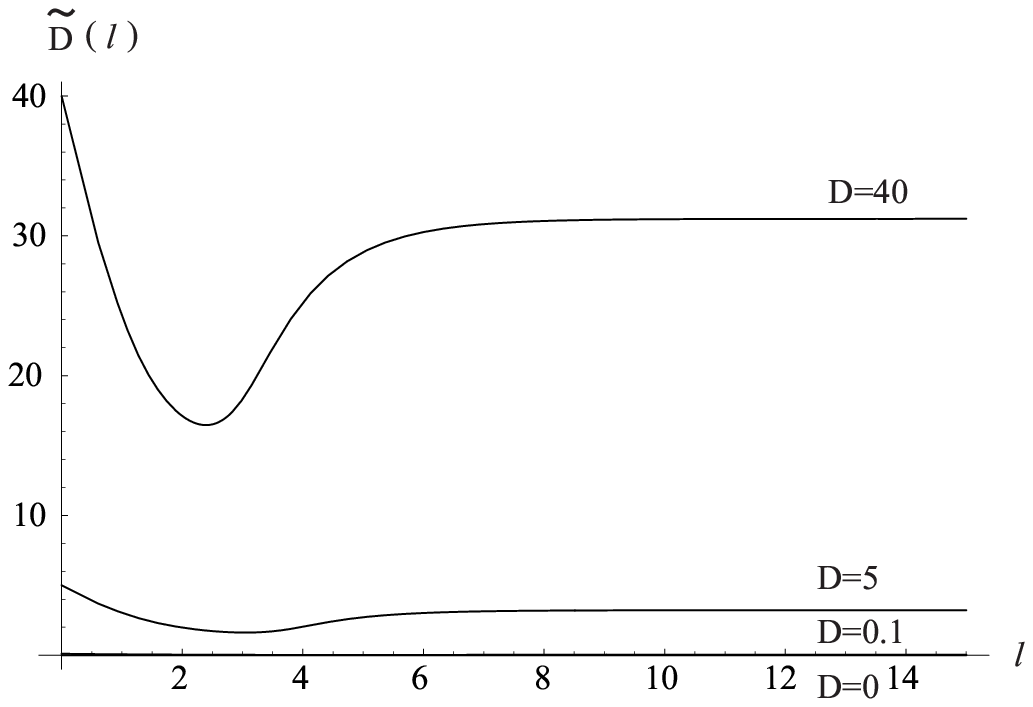}
\hspace{5mm}
\includegraphics[width=6cm,height=6cm,keepaspectratio,clip]{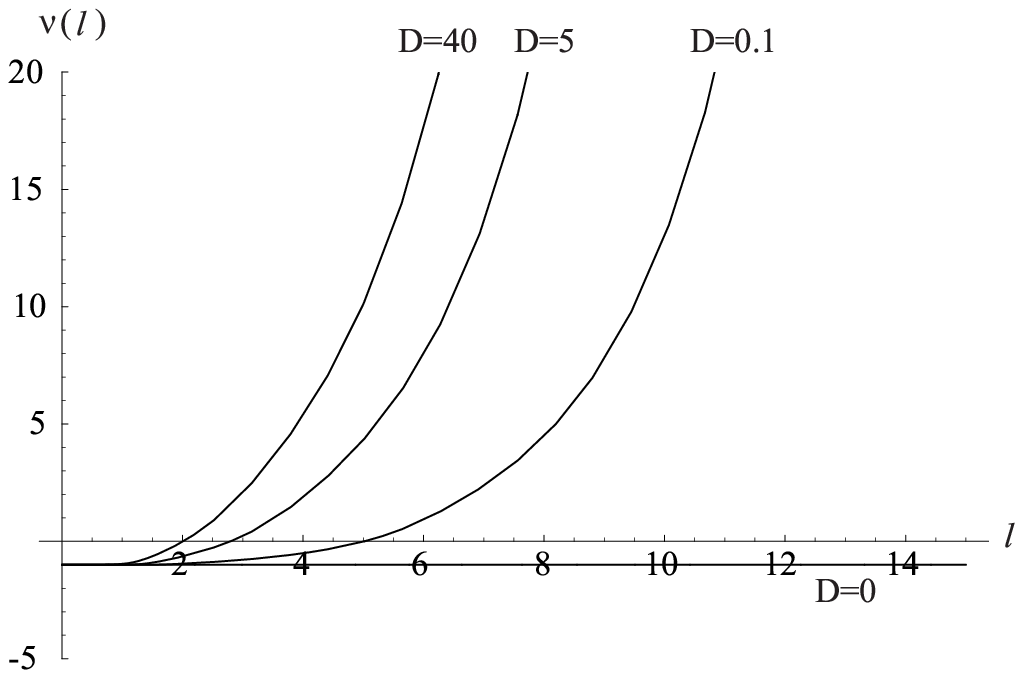}
\hspace{5mm}
\includegraphics[width=6cm,height=6cm,keepaspectratio,clip]{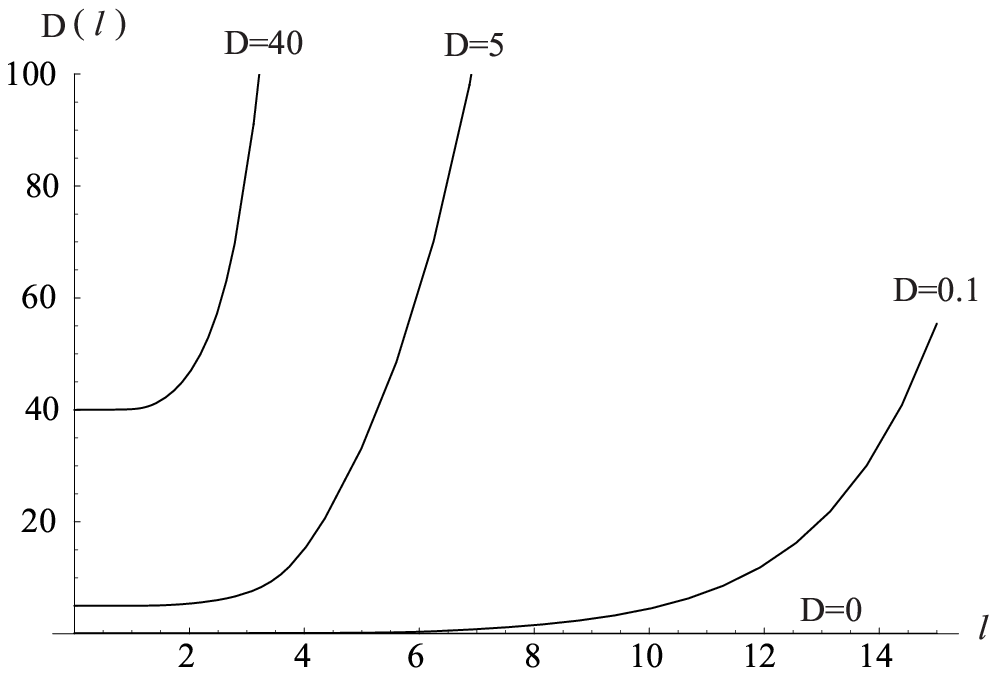}
\end{center}
\caption{(a) The rescaled viscosity $\tilde{\nu}(l)$, (b) the rescaled noise strength $\tilde{D}(l)$, (c) the viscosity $\nu(l)$ in the original scale, and (d) the noise strength $D(l)$ in the original scale, for $\tilde{\nu}(0)=\nu(0)=-1$, $\tilde{K}(0)=K(0)=1$, $\tilde{D}(0)=D(0)=D=0,0.1,5,40$, and $\tilde{D}_{d}(0)=D_{d}(0)=0$.}
\label{fig:nu-D-l}
\end{figure}

Figures \ref{fig:nu-D-l} (a) and (b) show the behavior of the rescaled viscosity $\tilde{\nu}(l)$ and noise strength $\tilde{D}(l)$, which are obtained by solving Eqs. (\ref{subeq:nu}) and (\ref{subeq:D}) by the use of Eqs. (\ref{subeq:F})-(\ref{subeq:H}) under the assumption that the values of $\alpha$ and $z$ are given by $\alpha=0.5$ and $z=1.5$. It is found that $\tilde{\nu}$ and $\tilde{D}$ converge to steady-state values $\tilde{\nu}^{*}$ and $\tilde{D}^{*}$. Their values become large with increase of $\tilde{D}(0)=D$. On the other hand, Figs. \ref{fig:nu-D-l} (c) and (d) show $\nu(l)$ and $D(l)$ in the original scale, which are obtained by solving Eqs. (\ref{subeq:orig-nu}) and (\ref{subeq:orig-D}) by the use of Eqs. (\ref{subeq:F})-(\ref{subeq:H}) or by making the scale transformation $\nu(l)=e^{-(z-2)l}\tilde{\nu}(l)$ and $D(l)=e^{-(z-2\alpha-1)l}\tilde{D}(l)$. It is found that $\nu(l)$ and $D(l)$ increase with the parameter $l$, which indicates that the RG analysis cannot predict the values of the effective viscosity $\nu_{\rm eff}$ and the effective noise strength $D_{\rm eff}$ in the intermediate scaling region where the linear term dominates the nonlinearity in the effective equation obtained from Eq. (\ref{eq:intro3}) \cite{Zaleski, Sneppen, Hayot, Sakaguchi}. By using a discrete stochastic model of erosion processes by ion sputtering, it was found that a periodic ripple morphology characterizes the initial stage of the evolution, whereas the surface displays self-affine scaling in the later stage \cite{Cuerno}. In the RG procedure one must rescale the space, time, and height in order to obtain a system similar to the original, which is based on the assumption that the interface is self-affine. Therefore we can say that the RG analysis can determine the self-affine scaling exponents of rough interfaces, but cannot predict the crossover from the ripple structure to the rough surface. In Figs. \ref{fig:nu-D-l} (a), (b), (c), and (d), we have chosen the initial values, $\tilde{\nu}(0)=\nu(0)=-1$, $\tilde{K}(0)=K(0)=1$, $\tilde{D}(0)=D(0)=D=0,0.1,5,40$, $\tilde{D}_{d}(0)=D_{d}(0)=0$, and $\Lambda_{0}=\pi/\Delta x=\pi/0.5$. It should be noted that the values of $(F^{*}, G^{*}, H^{*})$ are universal in the sense that they do not depend on the initial values, while $\tilde{\nu}^{*}$ and $\tilde{D}^{*}$ depend on them.

\subsection{Fluctuation-dissipation theorem}
From the linear part of the effective equation obtained from the noisy KS equation, the velocity correlation function in the original scale can be expressed as \cite{Forster, Yakhot-She}
\begin{equation}
C(k,\omega)
=\frac{\left<u(k, \omega)u(k', \omega')\right>}{(2\pi)^{2}\delta(k+k')\delta(\omega+\omega')}
=\frac{2\left(D(l)k^{2}+D_{d}(l)k^{4}\right)}
{\omega^{2}+\left(\nu(l)k^{2}+K(l)k^{4}\right)^{2}}.
\label{eq:corre1}
\end{equation}
On the other hand, the velocity correlation function in terms of the rescaled variables is given by
\begin{equation}
\tilde{C}(\tilde{k}, \tilde{\omega})
=\frac{2\left(\tilde{D}(l)\tilde{k}^{2}+\tilde{D}_{d}(l)\tilde{k}^{4}\right)}
{\tilde{\omega}^{2}
+\left(\tilde{\nu}(l)\tilde{k}^{2}+\tilde{K}(l)\tilde{k}^{4}\right)^{2}}.
\label{eq:corre2}
\end{equation}
$C(k,\omega)$ and $\tilde{C}(\tilde{k}, \tilde{\omega})$ are related as $\tilde{C}(\tilde{k}, \tilde{\omega})=e^{(1-2\alpha-z)l}C(k,\omega)$ by the scale transformation, then the scaling solution is given by $C(k,\omega)=k^{1-2\alpha-z}\Psi(\omega/k^{z})$, where $\Psi(x)$ is a scaling function. Substituting the values $z=1.5$ and $\alpha=0.5$, we obtain $C(k, \omega)=k^{-1.5}\Psi\left(\omega/k^{1.5}\right)$. Thus, the scaling solution for the effective equation obtained from the noisy KS equation is the same as that for the KPZ equation \cite{Kardar} and the KS equation \cite{L'vov, Fujisaka} in 1+1 dimensions. 
It follows from (\ref{eq:corre1}) that the energy spectrum of velocity in the original scale can be written as 
\begin{equation}
E(k)=\int \frac{d\omega}{2\pi}C(k, \omega)
=\frac{D(l)+D_{d}(l)k^{2}}{\nu(l)+K(l)k^{2}}
=\frac{D(l)}{\nu(l)}
\frac{1+\frac{H(l)}{G(l)}\left(\frac{k}{\Lambda(l)}\right)^{2}}
{1+\frac{1}{F(l)}\left(\frac{k}{\Lambda(l)}\right)^{2}},
\label{eq:corre3}
\end{equation}
while the energy spectrum of velocity in terms of the rescaled variables is given by
\begin{equation}
\tilde{E}(\tilde{k})=\frac{\tilde{D}(l)+\tilde{D}_{d}(l)\tilde{k}^{2}}{\tilde{\nu}(l)+\tilde{K}(l)\tilde{k}^{2}}
=\frac{\tilde{D}(l)}{\tilde{\nu}(l)}
\frac{1+\frac{H(l)}{G(l)}\left(\frac{\tilde{k}}{\Lambda_{0}}\right)^{2}}
{1+\frac{1}{F(l)}\left(\frac{\tilde{k}}{\Lambda_{0}}\right)^{2}}.
\label{eq:corre4}
\end{equation}
$E(k)$ and $\tilde{E}(\tilde{k})$ are related as $\tilde{E}(\tilde{k})=e^{(1-2\alpha)l}E(k)$ by the scale transformation, then the scaling solution is given by $E(k) \propto k^{1-2\alpha}$. At the fixed point, the values $(F^{*}, G^{*}, H^{*})=(10.7593, 680.652, 63.2614)$ yield the estimates $H^{*}/G^{*}\approx 0.093$ and $1/F^{*}\approx 0.093$. Therefore $\tilde{E}(\tilde{k})$ at the fixed point is $\tilde{D}^{*}/\tilde{\nu}^{*}$. 
Using the values $z=1.5$ and $\alpha=0.5$, we obtain $\nu(l)=e^{0.5(l-l_{0})}\tilde{\nu}^{*}$ and $D(l)=e^{0.5(l-l_{0})}\tilde{D}^{*}$, where we have used the fact that for $l_{0} \leq l$, $\tilde{\nu}(l)$ and $\tilde{D}(l)$ take the steady-state values $\tilde{\nu}^{*}$ and $\tilde{D}^{*}$ as shown in Figs. \ref{fig:nu-D-l} (a) and (b). Using the cutoff $\Lambda(l)=\Lambda_{0}e^{-l}$, they can be expressed as $\nu(l)=\left(\Lambda_{0}/\Lambda(l)\right)^{0.5(1-l_{0}/l)}\tilde{\nu}^{*}$ and $D(l)=\left(\Lambda_{0}/\Lambda(l)\right)^{0.5(1-l_{0}/l)}\tilde{D}^{*}$. Therefore in the limit $l \rightarrow \infty$, $\nu(l) \sim \Lambda(l)^{-0.5}$ and $D(l) \sim \Lambda(l)^{-0.5}$. This shows that the values of $\nu$ and $D$ in the original scale increase with decreasing the cutoff $\Lambda(l)$ \cite{Zaleski,com-L'vov}. However, their ratio remains unchanged, $D(l)/\nu(l)=\tilde{D}^{*}/\tilde{\nu}^{*}$, which is found to be a scale independent quantity. 
Accordingly, $E(k)$ also can be put as $\tilde{D}^{*}/\tilde{\nu}^{*}$, which indicates that in 1+1 dimensions the fluctuation-dissipation theorem holds in the noisy KS equation too in the long-wavelength limit. If we do not introduce $D_{d}$ term in the noise correlator, $E(k)$ without $H(l)$ in Eq. (\ref{eq:corre3}) depends on $k$ because of $1/F(l) \neq 0$. In the numerical simulations of the KS equation without noise term, the equal-time correlation function takes the steady-state value for $k \rightarrow 0$ \cite{Zaleski, Sneppen, Hayot}. The present RG analysis has shown that this is also the case in the noisy KS equation. 

\section{numerical analysis}
\subsection{Scaling exponents}
The long-wavelength properties of the KS equation are expected to behave like the KPZ equation. Sneppen $\etal$ performed a large-scale numerical simulation of the KS equation without the noise term \cite{Sneppen}. They investigated the dynamic scaling of $\langle (h-\langle h\rangle)^{2}\rangle^{1/2}\sim t^{\beta}$ and found crossover from the Edward-Wilkinson scaling with $\beta=1/4$. However, they could not find the KPZ scaling with $\beta=1/3$ clearly.  Much larger scale and longer time numerical simulations may be necessary to confirm the KPZ scaling. The RG analysis in the previous section shows that the fixed point $(F^{*},G^{*},H^{*})$ is more easily attained by stochastic noises. We expect therefore that the KPZ scaling with $\beta=1/3$ can be numerically observed in the noisy KS equation with moderate system  size. 

In the usual numerical simulations of the KS equation without the noise term \cite{Zaleski, Sneppen, Hayot, Sakaguchi}, all the parameters in Eq. (\ref{eq:intro1}) or (\ref{eq:intro3}) can be fixed to 1 by appropriate rescaling. The only control parameter is then the system size $L$ \cite{Zaleski, Sneppen, Hayot, Sakaguchi}. On the other hand, in the noisy version of the KS equation, $D$ and $D_{d}$ are added as the control parameters. We have performed direct numerical simulations of Eq. (\ref{eq:intro1}) with $\nu=-1,\,K=1,\,\lambda=1$, and investigated the time evolution of $W(t)=\langle \left[h(x,t)-\langle h(x,t)\rangle\right]^2\rangle^{1/2}$. We have investigated both cases of $D\ne 0,\,D_d=0$ and $D=0,\,D_d\ne 0$, and obtained similar results. We will show only the results for $D\ne 0$ and $D_d=0$, since the case is more natural for the noisy surface growth \cite{Cuerno}. We have used the Heun method for the numerical simulation with $\Delta x=0.5$ and $\Delta t=0.005$. (Sneppen $\etal$ used the Euler method with $\Delta x=1$ and $\Delta t=0.1$, which might be too rough discretization.) The periodic boundary conditions are imposed, and the initial condition is $h(x,0)=0$. 

\begin{figure}
\begin{center}
\includegraphics[width=6cm,height=6cm,keepaspectratio,clip]{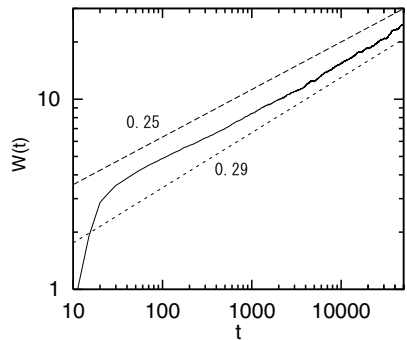}
\end{center}
\caption{Time evolution of $W(t)$ for the KS equation without the noise term for $L=200,000$. Two straight dashed lines denote the line with exponent $0.25$ and $0.29$.}
\label{fig:width-t(KS)}
\end{figure}

Figure \ref{fig:width-t(KS)} displays the time evolution of $W(t)$ for the KS equation for $L=200,000$. The ensemble average is taken for three runs. In this simulation, the noise term with $D=0.1$ is added only for the initial time interval $0<t<0.5$, and the time evolution obeys the deterministic KS equation for $t>0.5$. 
The double-logarithmic plot of $W(t)$ shows that $W(t)$ obeys the dynamic scaling $t^{1/4}$ fairly well. The slightly upward-curving line of $W(t)$ probably represents the crossover to the KPZ scaling. The exponent near $t\sim 50,000$ is about 0.29, which is still rather smaller than 1/3. That is, we could not confirm the KPZ scaling with $\beta=1/3$ clearly, either. 

\begin{figure}
\begin{center}
\includegraphics[width=6cm,height=6cm,keepaspectratio,clip]{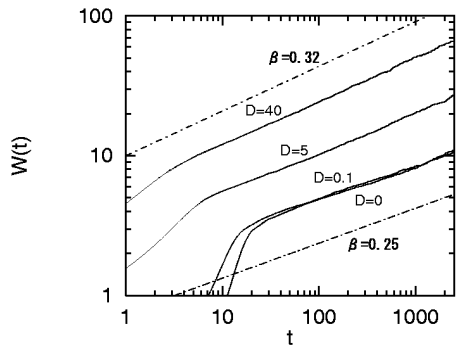}
\end{center}
\caption{Time evolutions of $W(t)$ for the noisy KS equation at $D=0,\,0.1,\,5$, and 40 for $L=20,000$.}
\label{fig:width-t(nKS)}
\end{figure}

Figure \ref{fig:width-t(nKS)} displays the time evolution of $W(t)$ for the noisy KS equation with $D=0,\,0.1,\,5,\,40$ for $L=20,000$. The ensemble average is taken for 10 runs. 
As the noise strength is increased, the time evolutions of $W(t)$ are shifted upwards, since the fluctuations increase owing to the stochastic noises.
For the deterministic case of $D=0$, only the dynamic scaling with $1/4$ is observed and even the crossover is hardly seen in this smaller size and shorter time simulation. For $D=0.1$, the time evolution is almost the same as the case of $D=0$. For $D=5$, the exponent of the dynamic scaling increases from $\beta=1/4$ towards $\beta=0.3$. For $D=40$, the exponent of the dynamic scaling is almost 0.32, which is close to the exponent 1/3 of the KPZ scaling. This result is consistent with the RG result in Fig. \ref{fig:F-G,F-H}, which shows that the KPZ fixed point is easily attained by the stochastic noises. If we can perform further extensive numerical simulation than that shown in Fig. \ref{fig:width-t(KS)}, the RG trajectory for the KS equation (i.e., the case of $D=D_{d}=0$) would reach the KPZ fixed point. When $D=D_{d}=0$, however, $\Sigma$ in Eq. (\ref{eq:Fs5}) and $\Phi$ in Eq. (\ref{eq:Fs6}) are always zero. Then the RG trajectory is on the $F$ axis in Fig \ref{fig:F-G,F-H} and never reach the KPZ fixed point. This is a qualitative difference between the RG analysis in the previous section and the numerical simulation. This is probably because the chaotic fluctuations generated by the deterministic KS equation are not well involved in the RG analysis when $D=D_{d}=0$. 

\subsection{Modeling by the KPZ equation and estimate of the effective parameters}
Next, we analyze the numerical results of the noisy KS equation, based on the KPZ equation with the effective noise strength and the effective viscosity, assuming that the modeling by the effective KPZ equation is a good approximation for the behavior in the large spatial and temporal scales for the case of small $D$. The KPZ equation is written as 
\begin{equation}
h_{t}=\nu_{\rm eff} h_{xx}+\frac{\lambda}{2} (h_{x})^2+\eta(x,t),
\label{eq:Numeric3}
\end{equation}
\begin{equation}
\left<\eta(x,t)\eta(x',t')\right>=2D_{\rm eff}\delta(x-x')\delta(t-t').
\label{eq:Numeric4}
\end{equation}
Sneppen $\etal$ evaluated the effective viscosity constant $\nu_{\rm eff}$ and the effective noise strength $D_{\rm eff}$ from the direct numerical simulations of the deterministic KS equation.  
They evaluated $D_{\rm eff}/\nu_{\rm eff}^{1/2}$ using the relation 
\begin{equation}
W(t)^2=\frac{2D_{\rm eff}}{\sqrt{2\pi\nu_{\rm eff}}}t^{1/2},
\label{eq:Numeric1}
\end{equation}
in the temporal range of the dynamic scaling with $\beta=1/4$. In the long-wavelength region, the equilibriumlike equipartition law for $h_{x}$ is observed for the KS equation \cite{Zaleski, Sneppen, Hayot, Sakaguchi}. Then, the second relation 
\begin{equation}
\frac{D_{\rm eff}}{\nu_{\rm eff}}=Lk^2\langle |h(k,t)|^2\rangle
\label{eq:Numeric2}
\end{equation}
determines the ratio of $D_{\rm eff}$ and $\nu_{\rm eff}$. From the two relations, Sneppen $\etal$ evaluated $D_{\rm eff}$ and $\nu_{\rm eff}$ as $D_{\rm eff}=6.4$ and $\nu_{\rm eff}=10.5$. They performed numerical simulation of the KS equation (\ref{eq:intro1}) with $\nu=-1,\,K=1,\,\lambda=2$, and the above value of $D_{\rm eff}$ is a rescaled value for $\lambda=1$. We note that the results by them are obtained by the replacements $h \rightarrow 2h$, $\eta \rightarrow 2\eta$, and $D_{\rm eff} \rightarrow 2D_{\rm eff}$ from the present results.
The effective viscosity $\nu_{\rm eff}$ was estimated with other methods \cite{Zaleski, Sakaguchi}, and similar values of $6<\nu_{\rm eff}<10$ were obtained. 
We have evaluated the effective viscosity $\nu_{\rm eff}$ and the effective noise strength $D_{\rm eff}$ for the noisy KS equation using the two relations (\ref{eq:Numeric1}) and (\ref{eq:Numeric2}). We have performed numerical simulations of $L=20,000$ to evaluate the relation (\ref{eq:Numeric1}), and numerical simulations of $L=4096$ to evaluate the relation (\ref{eq:Numeric2}). 

\begin{figure}
\begin{center}
\includegraphics[width=15cm,height=6cm,keepaspectratio,clip]{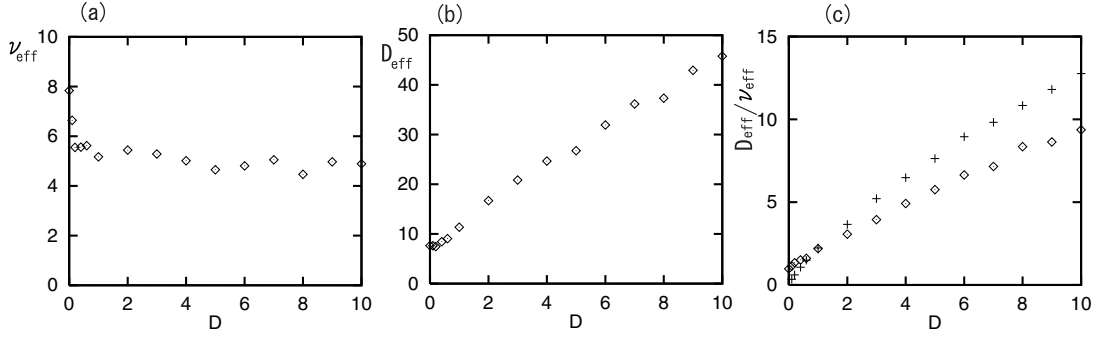}
\end{center}
\caption{(a) Effective viscosity and (b) the effective noise strength numerically estimated using the two relations (\ref{eq:Numeric1}) and (\ref{eq:Numeric2}) as a function of $D$. (c) The ratio $D_{\rm eff}/\nu_{\rm eff}$ obtained by numerical simulations (denoted by rhombi) and the RG analysis (denoted by crosses).}\label{fig:nu-D}
\end{figure}

Figure \ref{fig:nu-D} (a) and (b) display $\nu_{\rm eff}$ and $D_{\rm eff}$ as a function of $D$.
The effective viscosity $\nu_{\rm eff}$ decreases from $\nu_{\rm eff}\sim 8$ near $D=0$, and takes a value of about 5 for large $D$. The effective noise strength $D_{\rm eff}$ increases with $D$, since the stochastic noises are added to the deterministic noises generated spontaneously by the chaotic behavior. The values of $\nu_{\rm eff}$ and $D_{\rm eff}$ at $D=0$ are slightly different from the values estimated by Sneppen $\etal$ Figure \ref{fig:nu-D}(c) displays the ratio $D_{\rm eff}/\nu_{\rm eff}$ as a function of $D$. The rhombi denote the numerical values determined by Eq. (\ref{eq:Numeric2}) and the crosses denote the ratio $\tilde{D}/\tilde{\nu}$ obtained by solving RG flow equations (\ref{subeq:nu}) and (\ref{subeq:D}) with $\Lambda_{0}=\pi/\Delta x=\pi/0.5$, assuming that the values of $\alpha$ and $z$ are given by $\alpha=0.5$ and $z=1.5$, and the initial values are $\tilde{\nu}(0)=\nu=-1,\,\tilde{K}(0)=K=1,\,\tilde{D}(0)=D$, and $\tilde{D}_{d}(0)=D_{d}=0$. As we have shown in Fig \ref{fig:nu-D-l} (a) and (b), $\tilde{\nu}$ and $\tilde{D}$ approach steady-state values $\tilde{\nu}^{*}$ and $\tilde{D}^{*}$. We have evaluated the ratio $\tilde{D}/\tilde{\nu}$ at the steady-state. Even if returning to the original scale, the ratio remains unchanged, $D_{\rm eff}/\nu_{\rm eff}=\tilde{D}^{*}/\tilde{\nu}^{*}$, as discussed in the previous section. It may be said that the RG analysis gives roughly approximate values of $D_{\rm eff}/\nu_{\rm eff}$ for the long-wavelength fluctuations except for the range of small $D$.  

\subsection{Approximation of the surface growth law using the RG analysis}

The increase of the effective noise strength is expected to facilitate the crossover to the KPZ scaling as we have shown in Fig. \ref{fig:width-t(nKS)}. 
We consider the dynamic scaling based on the KPZ equation. We assume that $\nu_{\rm eff}$ and $D_{\rm eff}$ are evaluated using Eqs. (\ref{eq:Numeric1}) and (\ref{eq:Numeric2}).
The RG flow equation for $g(l)=\tilde{\lambda}(l)^2\tilde{D}_{\rm eff}(l)/(4\pi\tilde{\nu}_{\rm eff}(l)^3\Lambda_{0})$ is given as
\begin{equation}
\frac{dg(l)}{dl}=g(l)-2g(l)^2,
\label{eq:Numeric5}
\end{equation}
which can be obtained also from Eqs. (\ref{subeq:nu}), (\ref{subeq:lambda}), and  (\ref{subeq:D}) by putting $\tilde{K}=\tilde{D}_d=0$.
Equation (\ref{eq:Numeric5}) is easily solved as $g(l)=\{1+\tanh(l/2-c)\}/4$, where $c$ is determined from $g(0)=\lambda^2D_{\rm eff}/(4\pi\nu_{\rm eff}^3\Lambda_{0})=(1-\tanh c)/4$ with the use of the initial values of $\tilde{\nu}_{\rm eff}(0)=\nu_{\rm eff}$, $\tilde{D}_{\rm eff}(0)=D_{\rm eff}$, and $\tilde{\lambda}(0)=\lambda$. 
If we choose $g(0)=1/2$, we obtain $c=-\infty$. The substitution of $c=-\infty$ in the above exact solution of $g(l)$ yields $g(l)=1/2$ for any $l$. That is, if we choose the initial values of $\nu_{\rm eff}$, $D_{\rm eff}$, $\lambda$, and $\Lambda_0$ to satisfy $g(0)= \lambda^2D_{\rm eff}/(4\pi\nu_{\rm eff}^3\Lambda_{0})=1/2$, the dynamic scaling with $\beta=1/3$ is easily observed even for small $t$, since the system is at the fixed point of the RG flow. 
The rescaled parameters $\tilde{\nu}_{\rm eff}(l)$ and $\tilde{D}_{\rm eff}(l)$ obey Eqs. (\ref{subeq:nu}) and  (\ref{subeq:D}) by putting $\tilde{K}=\tilde{D}_d=0$. Returning to the original scale, the parameters $\nu_{\rm eff}(l)=e^{-(z-2)l}\tilde{\nu}_{\rm eff}(l)$ and $D_{\rm eff}(l)=e^{-(z-2\alpha-1)l}\tilde{D}_{\rm eff}(l)$ obey the equations
\begin{eqnarray}
\frac{d\nu_{\rm eff}(l)}{dl}&=&\nu_{\rm eff}(l)g(l),\nonumber\\
\frac{dD_{\rm eff}(l)}{dl}&=&D_{\rm eff}(l)g(l).
\label{eq:Numeric6}
\end{eqnarray}
The effective viscosity $\nu_{\rm eff}(l)$ and $D_{\rm eff}(l)$ increase as $\nu_{\rm eff}(l)=\nu_{\rm eff}\exp\{\int_0^lg(l)dl\}=\nu_{\rm eff}\exp[\{l+2\log \cosh(l/2-c)\}/4]$ and $D_{\rm eff}(l)=D_{\rm eff}\exp[\{l+2\log \cosh(l/2-c)\}/4]$.   
For $l\rightarrow \infty$, the wave number $k$ is scaled as $k\sim \Lambda_{0} e^{-l}$, so $l\sim \log (\Lambda_{0}/k)$. That is, the effective viscosity and noise strength in the original scale increase as $\nu_{\rm eff}(k)\sim k^{-1/2}$ and $D_{\rm eff}(k)\sim k^{-1/2}$ for small $k$. If we assume that $\nu_{\rm eff}(l)$ and $D_{\rm eff}(l)$ are the effective viscosity and the noise strength at the wave number $k=\Lambda_{0} e^{-l}$, the time evolution of $W(t)$ is expected to obey approximately 
\begin{equation}
W(t)^2=\sum_{k}\langle |h(k, t)|^2\rangle
\sim \sum_{k}\int_0^te^{-2\nu_{\rm eff}(k)k^2(t-t^{\prime})}2D_{\rm eff}(k)dt^{\prime}
=\sum_{k} \frac{D_{\rm eff}(k)}{\nu_{\rm eff}(k)k^{2}}
\left(1-e^{-2\nu_{\rm eff}(k)k^2t}\right),
\label{eq:W}
\end{equation}
where the summation is taken for the discrete wave numbers $k=k_n=2\pi n/L$ ($n=1, \cdots ,N/2$).
We can evaluate the time evolution of $W(t)$ using Eq.~(\ref{eq:W}). 

\begin{figure}
\begin{center}
\includegraphics[width=15cm,height=6cm,keepaspectratio,clip]{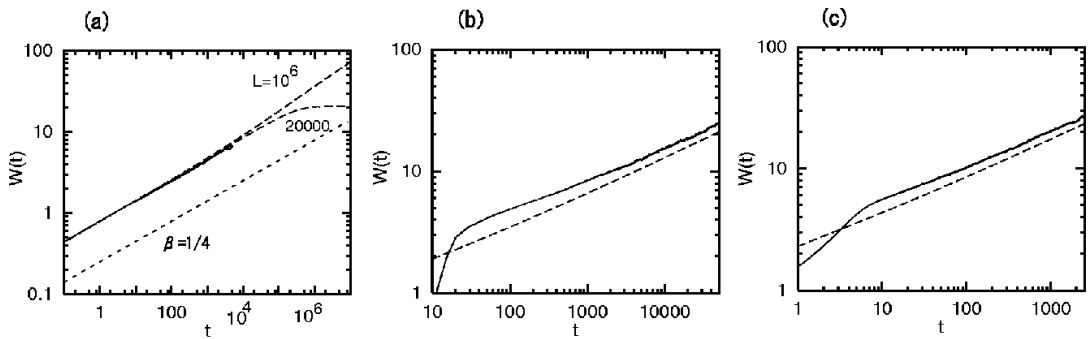}
\end{center}
\caption{Comparisons of $W(t)$ by direct numerical simulations (solid line) with Eq.~(\ref{eq:W}) (dashed lines) for (a) the KPZ equation with $\nu_{\rm eff}=10,\,D_{\rm eff}=5$ and $L=20,000$, (b) the KS equation for $L=200,000$ and (c) the noisy KS equation for $D=5$ and $L=20,000$.}
\label{fig:W-KPZ}
\end{figure}

Figure \ref{fig:W-KPZ} (a) compares the numerical result of the KPZ equation (\ref{eq:Numeric3}) for $L=20,000$, $\nu_{\rm eff}=10$, and $D_{\rm eff}=5$ with Eq.~(\ref{eq:W}) for $L=20,000$ and $1,000,000$. Equation~(\ref{eq:W}) gives a fairly good estimate for $W(t)$. The dynamic scaling with exponent $\beta=1/4$ is seen even for the KPZ equation at $\nu_{\rm eff}=10$. The saturation effect of $W(t)$ owing to the finiteness of $L$ is also expressed by Eq.~(\ref{eq:W}).
Figure \ref{fig:W-KPZ} (a) also shows that very large size and long time are necessary to see the crossover to the KPZ scaling for the large viscosity $\nu_{\rm eff}=10$. 
Figure \ref{fig:W-KPZ} (b) compares the numerical result of the KS equation at $D=0$ for $L=200,000$ (which is shown in Fig. \ref{fig:width-t(KS)}) with Eq. (\ref{eq:W}) shown by the dashed line for $L=200,000$ and $\nu_{\rm eff}=7.85,\,D_{\rm eff}=7.65$, where $\nu_{\rm eff}$ and $D_{\rm eff}$ are the values at $D=0$ in Fig. \ref{fig:nu-D} (a) and (b). Very slow crossover towards the KPZ scaling can be predicted by Eq.~(\ref{eq:W}). 
Figure \ref{fig:W-KPZ} (c) compares the numerical results of the noisy KS equation at $D=5$ for $L=20,000$ (which is shown in Fig. \ref{fig:width-t(nKS)}) with Eq.~(\ref{eq:W}) shown by the dashed line for $L=20,000$, $\nu_{\rm eff}=4.65,\,D_{\rm eff}=26.7$, where $\nu_{\rm eff}$ and $D_{\rm eff}$ are the values at $D=5$ in Fig. \ref{fig:nu-D} (a) and (b). The crossover towards the KPZ equation is clearly seen even in the smaller-size system. 

The corresponding summation formula for $W(t)$ using the RG flows of  Eqs.~(\ref{subeq:F})-(\ref{subeq:orig-Dd}) is written as
\begin{eqnarray}
W(t)^2
&\sim&
 \sum_{k}\int_{0}^{t}e^{-2\{\nu(k)k^2+K(k)k^4\}(t-t^{\prime})}
\{2D(k)+2D_d(k)k^2\}dt^{\prime}
\nonumber \\
&=&\sum_{k} \frac{D(k)+D_d(k)k^2}
{\nu(k)k^{2}+K(k)k^4}
\left(1-e^{-2\{\nu(k)k^2+K(k)k^4\}t}\right).
\label{eq:W2}
\end{eqnarray}
The growth law with exponent $\beta=1/4$ and the crossover towards the exponent $\beta=1/3$ did not appear in this formula. We suspect that fluctuations by the deterministic chaos are well involved in the RG analysis of Eqs.~(\ref{subeq:nu})-(\ref{subeq:orig-Dd}) especially for small $D$. 
That is a reason why we have assumed the KPZ equation for the basis of the dynamic scaling for small $D$. It might be related to a fact that 
the effective viscosities around $5\sim 10$ did not appear in the RG flows.

\begin{figure}
\begin{center}
\includegraphics[width=6cm,height=6cm,keepaspectratio,clip]{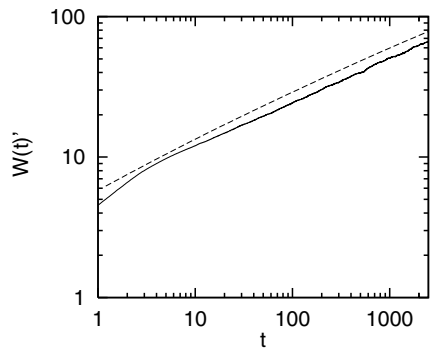}
\end{center}
\caption{Comparison of the time evolution of $W(t)$ for the nosiy KS equation with $D=40$ and $L=200,000$ (solid line) with Eq.~(\ref{eq:W2}) (dashed line).}
\label{fig:W-NKS}
\end{figure}

However, if $D$ is sufficiently large, the stochastic noises dominate the chaotic fluctuations, and the RG flows of Eqs.~(\ref{subeq:F})-(\ref{subeq:orig-Dd}) becomes more plausible. Besides, the effective viscosity and the effective noise strength cannot be evaluated for large $D$, since the intermediate region with the dynamic exponent $\beta=1/4$ becomes invisible, and the analyses based on the KPZ equation cannot be applied.  Here, we show the applicability of the RG flows for the noisy KS equation through Eq.~(\ref{eq:W2}) for large $D$. Figure \ref{fig:W-NKS} compares the numerical results for $L=20,000$ and $D=40$ shown in Fig. \ref{fig:width-t(nKS)} with Eq.~(\ref{eq:W2}). Fairly good agreement is seen for the large value of $D=40$. 

\section{conclusion}
We have performed the RG analysis for the KS equation with conserved and nonconserved noises in 1+1 dimensions. By introducing the conserved noise, first, we have obtained the values of the scaling exponents which are in good agreement with those at the KPZ fixed point. Second, we have shown that the fluctuation-dissipation theorem is exactly satisfied in the noisy KS equation in the long-wavelength limit. Therefore we can say that the long-wavelength properties of the noisy KS equation in 1+1 dimensions is described by the KPZ equation fairly well. However, the RG analysis for the noisy KS equation could not yet predict the effective viscosity and the noise strength in the intermediate scaling region. We have numerically evaluated the effective viscosity and the noise strength for the effective KPZ equation. We have shown that the KPZ scaling can be easily observed even in moderate-size numerical simulations of the KS equation under stochastic noises, owing to the increase of the effective noise strength. 

We have not succeeded in including the effect of the chaotic fluctuations into the RG flows and evaluating the effective viscosities for small $D$ or $D=0$. 
This problem also would arise in the other methods such as the direct interaction approximation (see Fig. 2a in \cite{Zaleski}) and the mode-coupling study \cite{Fujisaka} of the KS equation, which were unable to obtain spectra with a hump. A rough estimate of the effective viscosity is $\nu_{\rm eff} \approx l_{\rm cell}^{2}/\tau$, where $l_{\rm cell}$ is the size of the cellular structures in the KS system and $\tau$ is the linear growth time of the most unstable mode \cite{Sneppen, Hayot}. It is recognized that the range of linealy unstable modes, which are seen in the prominent hump in the spectra, play a crucial role in the nonlinear dynamical behaivor of the KS equation \cite{Procaccia}. Even if we are interested in the long-wavelength properties, we would not disregard this dynamically dominant range. In order to clarify whether the linearly unstable modes are incorporated by the RG method, it is worth to consider a model where the fourth-order derivative in Eqs. (\ref{eq:intro1}) or (\ref{eq:intro3}) is replaced by a sixth-order one. In such a variant model, it is reported that the values of $\nu_{\rm eff}$ and $D_{\rm eff}$ become huge, but the fluctuation-dissipation theorem is still well satisfied \cite{Zaleski, Hayot}. The huge value of $\nu_{\rm eff}$ can be qualitatively estimated from $\nu_{\rm eff} \approx l_{\rm cell}^{2}/\tau$. If the linearly unstable modes are well involved by the RG method, we would obtain the larger values of $\nu_{\rm eff}$ and $D_{\rm eff}$ for the variant model than those for the noisy KS equation. This is under investigation. 

\begin{acknowledgements}
We would like to thank H. Fujisaka and M. Kawasaki for useful comments.
\end{acknowledgements}


\begin{thebibliography}{99}

\bibitem{Kardar}
M. Kardar, G. Parisi, and Y.-C. Zhang, Phys. Rev. Lett. {\bf 56}, 889 (1986).

\bibitem{Kuramoto}
Y. Kuramoto and T. Tsuzuki, Prog. Theor. Phys. {\bf 55}, 356 (1976).

\bibitem{Sivashinsky}
G. I. Sivashinsky, Acta Astron. {\bf 4}, 1177 (1977).

\bibitem{Yakhot}
V. Yakhot, Phys. Rev. A {\bf 24}, R642 (1981).

\bibitem{Forster}
D. Forster, D. R. Nelson, and M. J. Stephen, Phys. Rev. A {\bf 16}, 732 (1977).

\bibitem{Yakhot-She}
V. Yakhot and Z.-S. She, Phys. Rev. Lett. {\bf 60}, 1840 (1988).

\bibitem{Zaleski}
S. Zaleski, Physica D {\bf 34}, 427 (1989).

\bibitem{Sneppen}
K. Sneppen, J. Krug, M. H. Jensen, C. Jayaprakash, and T. Bohr, Phys. Rev. A {\bf 46}, R7351 (1992).

\bibitem{Hayot}
F. Hayot, C. Jayaprakash, and C. Josserand, Phys. Rev. E {\bf 47}, 911 (1993).

\bibitem{Sakaguchi}
H. Sakaguchi, Prog. Theor. Phys. {\bf 107}, 879 (2002).

\bibitem{L'vov}
V. S. L'vov, V. V. Lebedev, M. Paton, and I. Procaccia, Nonlinearity {\bf 6}, 25 (1993).

\bibitem{Procaccia}
I. Procaccia, M. H. Jensen, V. S. L'vov, K. Sneppen, and R. Zeitak, Phys. Rev. A {\bf 46}, 3220 (1992).

\bibitem{Cuerno}
R. Cuerno, H. A. Makse, S. Tomassone, S. T. Harrington, and H. E. Stanley,  Phys. Rev. Lett. {\bf 75}, 4464 (1995).

\bibitem{Karma}
A. Karma and C. Misbah,  Phys. Rev. Lett. {\bf 71}, 3810 (1993).

\bibitem{R-Cuerno}
R. Cuerno and K. B. Lauritsen, Phys. Rev. E {\bf 52}, 4853 (1995).

\bibitem{Barabasi}
A.-L. Barab${\rm \acute{a}}$si and H. E. Stanley, {\it Fractal Concepts in Surface Growth} (Cambridge University Press, Cambridge, England, 1995).

\bibitem{Golubovic}
L. Golubovic and R. Bruinsma, Phys. Rev. Lett {\bf 66}, 321 (1991); {\bf 67}, 2747(E) (1991) .

\bibitem{Eyink}
G. L. Eyink, Phys. Fluids. {\bf 6}, 3063 (1994).

\bibitem{Fujisaka}
H. Fujisaka and T. Yamada, Prog. Theor. Phys. {\bf 57}, 734 (1977).

\bibitem{com-L'vov}
V. L'vov and I. Procaccia, Phys. Rev. Lett. {\bf 72}, 307 (1994).

\end{thebibliography}
\end{document}